\documentclass{aa}

\usepackage[varg]{txfonts}
\usepackage{graphicx}
\usepackage[colorlinks=true,citecolor=blue]{hyperref}
\usepackage{natbib}

\defcitealias{zhou2023}{Paper~I}   

\begin{document}

\title{Gas inflows from cloud to core scales in G332.83-0.55: Hierarchical hub-filament structures and tide-regulated gravitational collapse}

\author{J. W. Zhou\inst{\ref{inst1}} 
\and S. Dib\inst{\ref{inst2}}
\and M. Juvela\inst{\ref{inst3}}
\and P. Sanhueza\inst{\ref{inst4}}
\and F. Wyrowski\inst{\ref{inst1}}
\and T. Liu\inst{\ref{inst5}}
\and K. M. Menten\inst{\ref{inst1}}
}
\institute{Max-Planck-Institut f\"{u}r Radioastronomie, Auf dem H\"{u}gel 69, 53121 Bonn, Germany \label{inst1} \\
\email{jwzhou@mpifr-bonn.mpg.de}
\and
Max Planck Institute for Astronomy, K\"{o}nigstuhl 17, 69117 Heidelberg, Germany \label{inst2}
\and
Department of Physics, P.O.Box 64, FI-00014, University of Helsinki, Finland \label{inst3}
\and
National Astronomical Observatory of Japan, National Institutes of Natural Sciences, 2-21-1 Osawa, Mitaka, Tokyo 181-8588, Japan \label{inst4}
\and
Shanghai Astronomical Observatory, Chinese Academy of Sciences, 80 Nandan Road, Shanghai 200030, Peoples Republic of China \label{inst5}
}

\date{Accepted XXX. Received YYY; in original form ZZZ}

\abstract
{The massive star-forming region G332.83-0.55 contains at least two levels of hub-filament structures. The hub-filament structures may form through the "gravitational focusing" process.
High-resolution LAsMA and ALMA observations can directly trace the gas inflows from cloud to core scales. 
We investigated the effects of shear and tides from the protocluster on the surrounding local dense gas structures. Our results seem to deny the importance of shear and tides from the protocluster. However, for a gas structure, it bears the tidal interactions from all external material, not only the protocluster. To fully consider the tidal interactions, we derived the tide field according to the surface density distribution. Then, we used the average strength of the external tidal field of a structure to measure the total tidal interactions that are exerted on it. For comparison, we also adopted an original pixel-by-pixel computation to estimate the average tidal strength for each structure. Both methods give comparable results.
After considering the total tidal interactions, for the scaling relation between the velocity dispersion $\sigma$, the effective radius $R$, and the column density $N$ of all the structures, the slope of the $\sigma-N*R$ relation changes from 0.20$\pm$0.04 to 0.52$\pm$0.03, close to 0.5 of the pure free-fall gravitational collapse, and the correlation also becomes stronger. Thus, the deformation due to the external tides can effectively slow down the pure free-fall gravitational collapse of gas structures. The external tide tries to tear up the structure, but the external pressure on the structure prevents this process. The counterbalance between the external tide and external pressure hinders the free-fall gravitational collapse of the structure, which can also cause the pure free-fall gravitational collapse to be slowed down.
These mechanisms can be called "tide-regulated gravitational collapse."}


\titlerunning{Gas inflows from cloud to core scales in G332.83-0.55}
\authorrunning{J. W. Zhou}

\maketitle 

\section{Introduction}\label{intro}

High-resolution observations of high-mass star-forming regions show that density enhancements are organized in filamentary gas networks, particularly in hub-filament systems. 
Hub-filament systems  are known as a junction of three or more filaments. Filaments have lower column densities compared to the hubs \citep{Myers2009,Schneider2012,Kumar2020-642,Zhou2022-514}. In such systems, converging flows funnel matter into the hub through the filaments. Previous studies have suggested that hub-filament systems are birth cradles of high-mass stars and clusters 
\citep{Peretto2013,Henshaw2014,Zhang2015,Liu2016,Contreras2016-456,Yuan2018,Lu2018,Issac2019,Dewangan2020,Liu2021-646,Liu2022-511,Kumar2020-642,Zhou2022-514,Liu2023-522,Xu2023arXiv,Zhou2023-676,Yang2023-953}. 
\citet{Zhou2022-514}
studied the physical properties and evolution of hub–filament systems by spectral lines that were observed in the ATOMS (ALMA Three-millimeter Observations of Massive Star-forming regions) survey \citep{Liu2020}. We found that hub-filament structures can exist at various scales from 0.1 parsec to several parsec in very different Galactic environments. Below the 0.1 parcsec scale, slender structures similar to filaments, such as spiral arms, have also been detected in the surroundings of high-mass protostars \citep{Liu2015-804,Maud2017-467,Izquierdo2018-478,Chen2020-4,Sanhueza2021-915,Olguin2023-959}. Therefore, a scenario was proposed in \citet{Zhou2022-514}:
Self-similar hub-filament structures and filamentary accretion seem to exist at all scales (from several thousand au to several parsec) in high-mass star-forming regions. This paradigm of hierarchical and multi-scale hub-filament structures was generalized from  clump-core scale to cloud-clump scale in \citet{Zhou2023-676}. 
Hierarchical collapse and hub-filament structures feeding the central regions are also  described in previous works, such as \citet{Motte2018-56}, \citet{Vazquez2019-490}, \citet{Kumar2020-642} and references therein.

As discussed in \citet{Zhou2023-676},
the change in velocity gradients with the scale in the G333 complex indicates that the morphology of the velocity field in PPV space resembles a "funnel" structure. The funnel structure can be explained as accelerated material flowing toward the central hub and gravitational contraction of star-forming clouds or clumps. To some extent, the funnel structure gives an indication of the gravitational potential well that is formed by the clustering.

In this work, we focus on a local region G333-s3a in the G333 complex, called G332.83-0.55 hereinafter, which is marked in Fig.1 of \citet{Zhou2023-676}, also shown in Fig.\ref{map}. 
G332.83-0.55 presents a typical hub-filament morphology at $\sim$10 pc. As a cloud-scale hub-filament system, its hub corresponds to the ATLASGAL clump AGAL332.826-00.549. 
High-resolution APEX/LAsMA $^{12}$CO and $^{13}$CO (3-2) observation allowed us to trace the gas kinematics and gas inflows from cloud to clump scales in G332.83-0.55. We also zoomed in on the hub region of G332.83-0.55 using high-resolution ALMA HCO$^{+}$ (1-0) and H$^{13}$CO$^{+}$ (1-0) observations to investigate gas motions from clump to core scales. By combing LAsMA and ALMA observations, we could directly recover the hierarchical or multi-scale hub-filament structures and gas inflows in G332.83-0.55. In addition, we were also able to characterize the details of the funnel structure at different scales. 

\section{Observations and data}\label{data}

\subsection{LAsMA data}
The seven pixel Large APEX sub-Millimeter Array (LAsMA) receiver of the APEX telescope was used to observe the $J=3-2$ transitions of $^{12}$CO  and $^{13}$CO in the G333 molecular cloud complex. 
A detailed description of data calibration, imaging, and product creation procedures is presented in \citet{Zhou2023-676}.
The final data cubes have a velocity resolution of 0.25 km s$^{-1}$ and an angular resolution of $19.5 \arcsec$ and the pixel size is $6 \arcsec$. The final noise levels of the $^{12}$CO (3--2) and $^{13}$CO (3--2) data cubes are $\sim$0.32 K and $\sim$0.46 K, respectively. 
In this paper, we only focus on the subregion G332.83-0.55.

\subsection{ALMA data}
We selected one source from the ATOMS survey (Project ID: 2019.1.00685.S; PI: Tie Liu). The details of the 12m array and 7m array ALMA observations were summarized in \citet{Liu2020,Liuh2021}. 
Calibration and imaging were carried out using the CASA software package version 5.6 \citep{McMullin2007}; more details can be found in \citet{Zhou2021-508}.
All images have been primary-beam corrected. The continuum image reaches a typical 1 $\sigma$ rms noise of $\sim$0.2 mJy in a synthesized beam FWHM size of $\sim2.2\arcsec$. 
In this work, we also use the H$^{13}$CO$^+$ J=1-0 and HCO$^+$ J=1-0 molecular lines. The final noise levels of the H$^{13}$CO$^+$ J=1-0 and HCO$^+$ J=1-0 data cubes are $\sim$0.008 Jy beam$^{-1}$ and $\sim$0.012 Jy beam$^{-1}$, respectively. 

\subsection{Archival data}
We also used ATLASGAL+Planck 870 $\mu$m data (ATLASGAL combined with Planck data), which are sensitive to a wide range of spatial scales at a resolution of $\sim21\arcsec$ \citep{Csengeri2016-585}. 

\section{Results}

\subsection{Connection of gas inflows in LAsMA and ALMA observations}

\begin{figure*}
\centering
\includegraphics[width=1\textwidth]{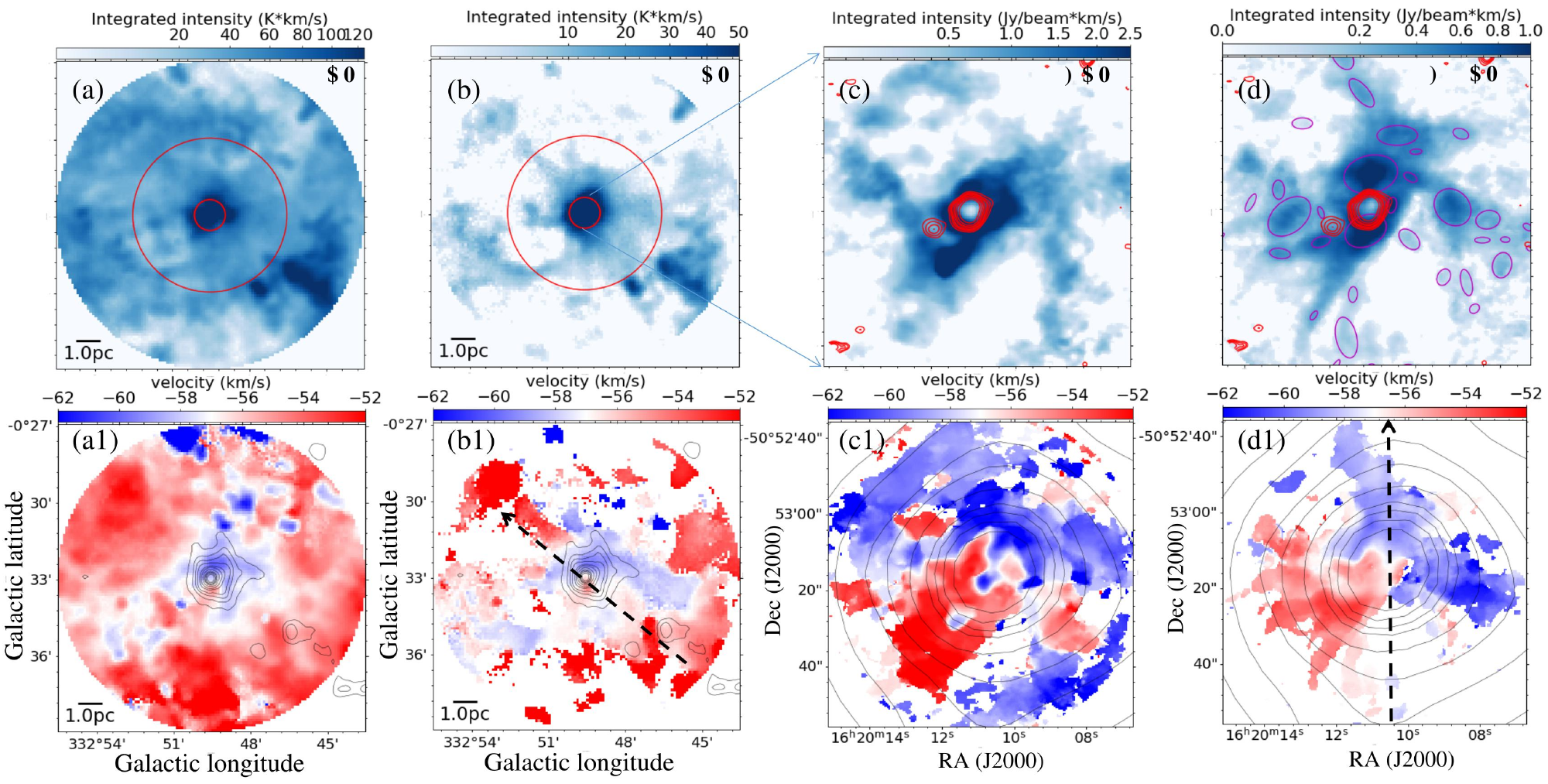}
\caption{First and second columns: Moment maps of LAsMA $^{12}$CO (3$-$2) and $^{13}$CO (3$-$2) emission. The radius of the entire region is 0.1$\degr$ ($\sim$6.28 pc).
The big red circle marks the region with a 0.05$\degr$ ($\sim$3.14 pc) radius, and the small red circle shows the region with a 0.01$\degr$ ($\sim$0.63 pc) radius. 
Third and fourth columns: Moment maps of ALMA 7m+12m HCO$^+$ J=1-0 and H$^{13}$CO$^+$ J=1-0 emission. The radius of the entire region is 0.01$\degr$ ($\sim$0.63 pc). Red contours in panels (c) and (d) show the 3mm continuum emission.
Black contours in the second row show ATLASGAL 870$\mu$m continuum emission.
Dashed black lines in panels (b1) and (d1) show the paths of the PV diagrams in Fig.\ref{pv}. Ellipses in panel (d) represent the dendrogram structures identified in Sec.\ref{dendro}.}
\label{map}
\end{figure*}

\begin{figure*}
\includegraphics[width=1\textwidth]{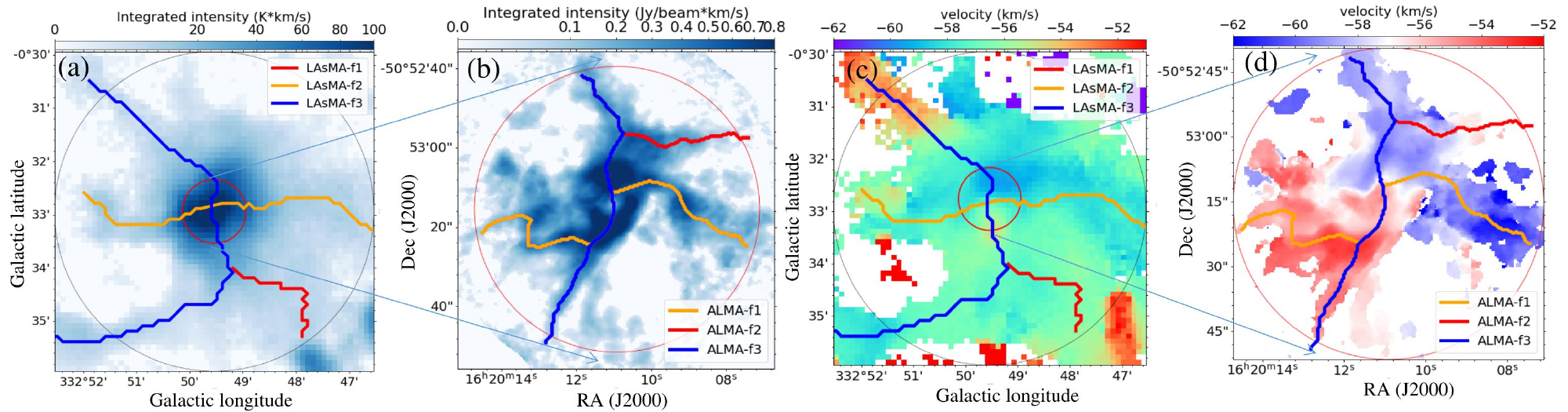}
\caption{ Hierarchical
hub-filament structures in G332.83-0.55. (a) and (b) Moment 0 maps of LAsMA $^{13}$CO (3$-$2) and ALMA H$^{13}$CO$^{+}$ (1-0) emission. (c) and (d)
Moment 1 maps of LAsMA $^{13}$CO (3$-$2) and ALMA H$^{13}$CO$^{+}$ (1-0) emission.
The hub region marked by a red circle in panels (a) and (c) are enlarged in panels (b) and (d).
More detailed structures can be seen in panels (b) and (d),
thanks to the high resolution of ALMA observation.
Lines in color present the filament skeletons identified by the FILFINDER algorithm. 
}
\label{flow1}
\end{figure*}

\begin{figure*}
\centering
\includegraphics[width=1\textwidth]{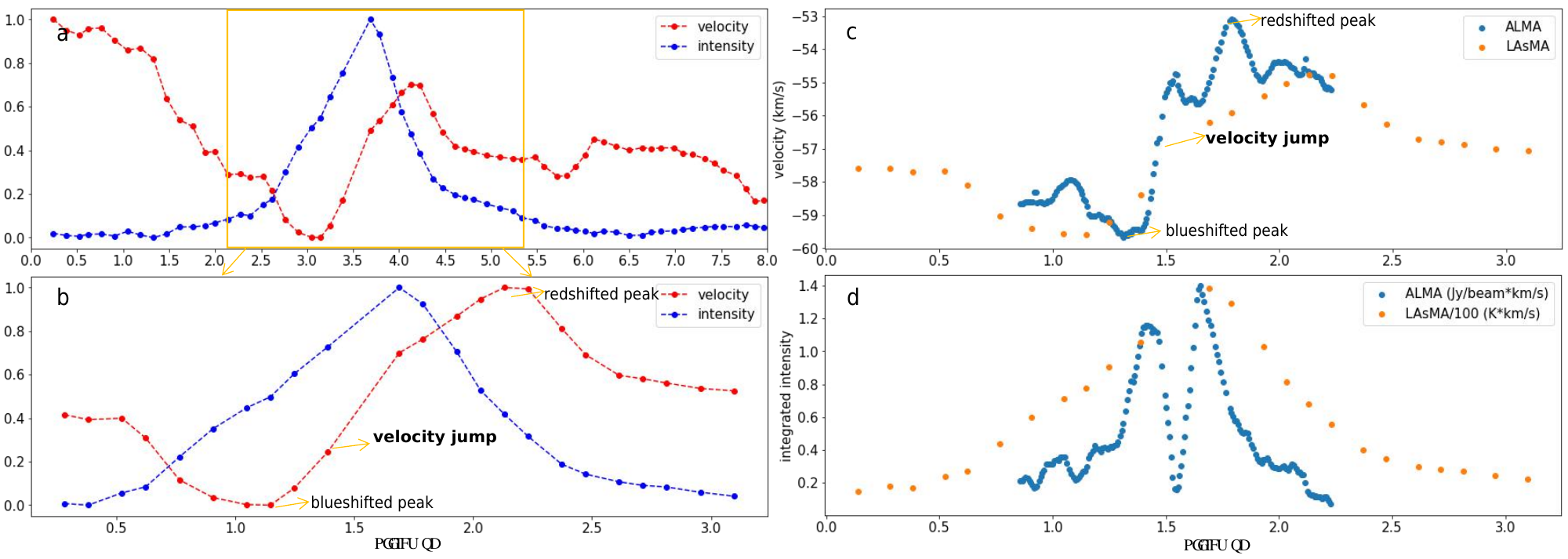}
\caption{(a) Normalized velocity and intensity along the filament LAsMA-f3 in Fig.\ref{flow1}(a); here we can see a blueshifted V-shape velocity structure. 
(b) Segment of panel (a) around the central hub; here we can see a V+$\Lambda$-shape velocity structure.
(c) and (d) Velocity and intensity along the segment of LAsMA-f3 in panel (b) compared to the velocity and intensity along the filament ALMA-f3 in Fig.\ref{flow1}(b).}
\label{flow}
\end{figure*}

\begin{figure}
\centering
\includegraphics[width=0.48\textwidth]{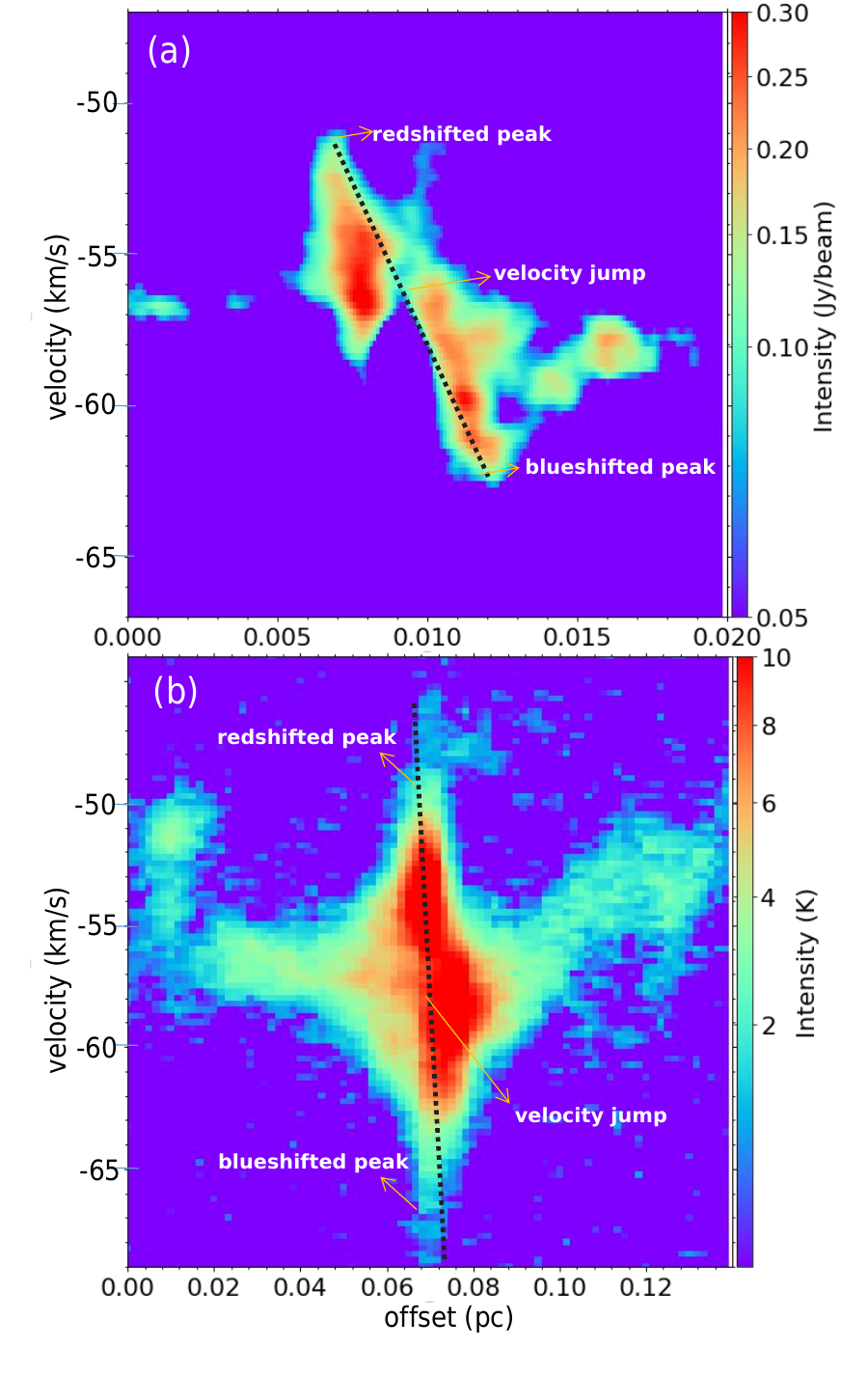}
\caption{PV diagrams of ALMA H$^{13}$CO$^{+}$ (1-0) and LAsMA $^{13}$CO (3$-$2) along the paths marked in Fig.\ref{map} 
(d1) and (b1).
}
\label{pv}
\end{figure}

\begin{figure*}
\centering
\includegraphics[width=1\textwidth]{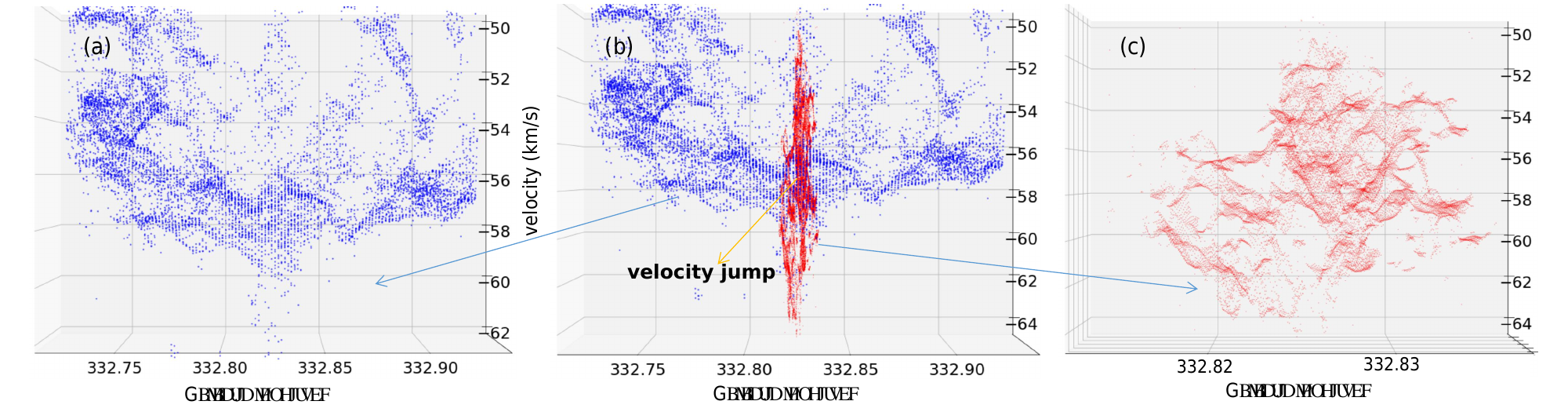}
\caption{Linkage of large- and small-scale velocity fields in PPV space. (a) Velocity field revealed by LAsMA data $^{13}$CO (3$-$2) on large scales decomposed by \texttt{GAUSSPY+}. (b) Combined velocity fields 
traced by $^{13}$CO (3$-$2) and H$^{13}$CO$^{+}$ (1-0) at different scales, i.e., panel (a) plus panel (c). 
(c) Velocity field revealed by ALMA data H$^{13}$CO$^{+}$ (1-0) on small scales decomposed by \texttt{GAUSSPY+}.}
\label{ppv}
\end{figure*}

Moment maps of various molecular lines are shown in Fig.\ref{map}.
The Moment 0 maps of $^{13}$CO (3$-$2), H$^{13}$CO$^+$ J=1-0, and HCO$^+$ J=1-0 emission show a typical hub-filament morphology. 
The FILFINDER algorithm \citep{Koch2015} was used to identify and characterize filaments traced by H$^{13}$CO$^{+}$ (1-0) and $^{13}$CO (3$-$2) emission. Following the method described in \citet{Zhou2022-514} and \citet{Zhou2023-676}, we used the integrated intensity (Moment 0) maps to identify filaments. The skeletons of the identified filaments, overlaid on the Moment 0 maps in Fig.\ref{flow1}, are highly consistent with the gas structures.  
We selected the filaments through the redshifted and blueshifted peaks (LAsMA-f3 and ALMA-f3), and extracted the velocity and intensity along the filaments, as shown in Fig.\ref{flow}.

Fig.\ref{pv} displays the position-velocity (PV) diagrams along the dashed lines in panels (b1) and (d1) of Fig.\ref{map}. For H$^{13}$CO$^+$ J=1-0 emission, feedback from the central UC-HII region identified in \citet{Zhang2023-520} has destroyed the surrounding molecular gas structure (Fig.\ref{map}(d)). Thus, the diagram looks broken, but it still has a clear double-peak velocity structure, indicating that the surrounding gas is still infalling onto the protocluster (Sec.\ref{infall}). In Fig.\ref{pv} (b), the central part of the PV diagram has a similar morphology to Fig.\ref{pv} (a). Large-scale gas structures with clear velocity gradients are connected to the central velocity jump. Correspondingly,
large-scale filamentary structures connected to the central clump can be seen in Fig.\ref{map}(a) and (b); small-scale filamentary structures down to dense core scales are also presented in Fig.\ref{map}(c) and (d). These features indicate that cloud-scale gas inflows extend down to core scales.
The inheritance of gas flows on large scales to small scales is clearly visible in Fig.\ref{flow}(c).

Fig.\ref{ppv}(a) displays the PPV map of $^{13}$CO (3$-$2) emission decomposed by the fully automated Gaussian decomposer  \texttt{GAUSSPY+} \citep{Lindner2015-149, Riener2019-628}; the decomposition was carried out in \citet{Zhou2023-676}. We can see a funnel structure that can be a probe of gravitational collapse as discussed in \citet{Zhou2023-676}.
At small scales, we can roughly identify a velocity jump comprised of the redshifted and blueshifted peaks in Fig.\ref{flow1}(c), which is clearly shown in Fig.\ref{flow1}(d), thanks to the higher resolution of the ALMA data. 

To conclude,
G332.83-0.55 shows at least two levels of hub-filament structures. Fig.\ref{map}(a) and (b) present the first level from cloud to clump scales. The second level from clump to core scales is shown in Fig.\ref{map}(c) and (d).
In a statistical study of hub-filament systems in \citet{Zhou2022-514}, we found filaments are ubiquitous in proto-clusters, and hub-filament systems are very common from dense core scales ($\sim$0.1 pc) to clump and cloud scales ($\sim$1-10 pc). These observed features are  reproduced in G332.83-0.55 well. 

The structures marked by small red circles in Fig.\ref{map}(a) and (b) present an approximately spherical morphology due to the limited resolution in single-dish observation, usually referred to as a "clump." However, in the eyes of ALMA, the spherical clump displays anisotropic and filamentary accretion structures (hub-filament) in Fig.\ref{map}(c) and (d). Thus, relative to the traditional "infall envelope," filamentary accretion and inflow structures may be more realistic in high-mass star-forming clumps.

\subsection{Infall+rotation motions}\label{infall}
In Fig.\ref{map}(c) and (d), 
on small scales, H$^{13}$CO$^+$ J=1-0 and HCO$^+$ J=1-0 emission present spiral-like structures, indicating the rotation of gas around the protocluster. PV diagrams in Fig.\ref{pv} also reveal rotation motions.
The PV diagrams show two peaks separated by several km s$^{-1}$, and they are symmetrically located with respect to the gravitational center. The double-peaked feature on the PV map is often seen in cluster-forming clumps, and this feature can only be reproduced when the clump has both infall motion and rotation at the same time, thus strongly implying that the clump is collapsing with rotation \citep{Ohashi1997-475,Shimoikura2016-832, Shimoikura2018-855, Sanna2019-623,Liu2020-904,Shimoikura2022-928}.  
Around the gravitational center, we can see the blueshifted and redshifted extrema at the end of the velocity jump in Fig.\ref{flow}, which is comparable with Fig.2 of \citet{Sakai2014-507}. In the schematic diagram of an infalling and rotating envelope model for IRAS 04368+2557 \citep{Sakai2014-507}, the shift of 
blueshifted and redshifted peaks can be used to estimate the radius of the centrifugal barrier $r_{\rm c}$, where the rotation velocity $V_{\rm rotation}$ has its maximum value.
According to equations (1) and (2) in \citet{Sakai2014-507}, we can estimate the mass of the gravitational center according to $r_{\rm c}$ by 
\begin{equation}
M_{*}= \frac{V_{\rm rotation}^{2}r_{\rm c}}{2G}.
\label{mass}
\end{equation}
The initial physical quantities in this equation can be estimated from the velocity jump in Fig.\ref{flow}. 

Both the velocity fields traced by LAsMA and ALMA data show a velocity jump around the gravitational center. 
In Fig.\ref{flow}(c), the velocity fields along the large- and small-scale filaments can be roughly connected, despite the significant differences in their resolutions (2.5$\arcsec$ and 19.5$\arcsec$). 
According to the velocity difference and the separation between blueshifted and redshifted peaks in Fig.\ref{flow}(c), we roughly estimate
$V_{\rm rotation,ALMA} \approx$3.29 km/s and $r_{\rm c, ALMA} \approx$ 0.21 pc. Then according to equation.\ref{mass}, the mass of the gravitational center (i.e., the protocluster) is $\sim$264 M$_{\odot}$.

As a further test, following the method described in \citet{Kauffmann2008}, the mass can be computed by dust emission using
\begin{eqnarray}
  M & = &
  \displaystyle 0.12 \, M_{\odot}
  \left( {\rm e}^{1.439 (\lambda / {\rm mm})^{-1}
      (T / {\rm 10 ~ K})^{-1}} - 1 \right) \nonumber \\
  & & \displaystyle
 \left( \frac{\kappa_{\nu}}{0.01 \rm ~ cm^2 ~ g^{-1}} \right)^{-1}
  \left( \frac{F_{\nu}}{\rm Jy} \right)
  \left( \frac{d}{\rm 100 ~ pc} \right)^2
  \left( \frac{\lambda}{\rm mm} \right)^{3},
  \label{mass_eq}
\end{eqnarray}
\noindent
where the opacity is 
\begin{equation}
\kappa_{\nu}=0.1({\nu}/\textrm{1000~GHz})^{\beta}~{\rm cm}^{2} {\rm g}^{-1}\end{equation}
and $\beta$ is the dust emissivity spectral index. Here, we consider $\beta=1.5$, 
$F_\nu$ is the total flux of 3mm dust emission, $d$ is the distance to the source, and $\lambda$ is the wavelength. The temperature $T$ takes the value $\sim$31.4 K listed in Table A1 of \citet{Liu2020}.
A mass
$\sim$264M$_{\odot}$ corresponds to the flux of 3mm dust emission
$\sim$0.12 Jy. This value is $\sim$3\% of the total flux of 3mm continuum emission in ALMA observations. As discussed in \citet{Keto2008-672} and \citet{Zhang2023-520},
3mm continuum emission of UC HII regions in high-resolution interferometric observations should be mostly free–free emission rather than from dust, and thus the proportion of $\sim$3\% for the dust emission is reasonable.


\subsection{Local dense gas structures}\label{dendro}
The issues of identifying the structures in a PPV cube by the dendrogram algorithm are described in \citet{Zhou2024-682}, and thus we adopted the same method as in \citet{Zhou2024-682} and \citet{Zhou2024-682-173} to identify dense gas structures. We directly identified hierarchical (sub)structures based on the 2D integrated intensity (Moment 0) map of H$^{13}$CO$^{+}$ (1$-$0) emission, and then extracted the average spectrum of each structure to investigate its velocity components and gas kinematics. More details can be found in \citet{Zhou2024-682-173}. 
In this work, we focus on leaf structures, which are local gravitational centers. According to their averaged spectra, the structures with absorption features were eliminated first. Following \citet{Zhou2024-682-173}, we only consider type1 (single velocity component) and type3 (blended velocity component) structures, as marked in Fig.\ref{map}(d) by ellipses.

We derived the column density by combining H$^{13}$CO$^+$ J=1-0 and HCO$^+$ J=1-0 emission using the same method described in Sec.3.3 of \citet{Zhou2024-682-173} by the local thermodynamic equilibrium (LTE) analysis. Then the masses and virial ratios of all leaf structures were calculated.


\subsection{Shear and tide from the protocluster}\label{tide1}

\begin{figure}
\centering
\includegraphics[width=0.48\textwidth]{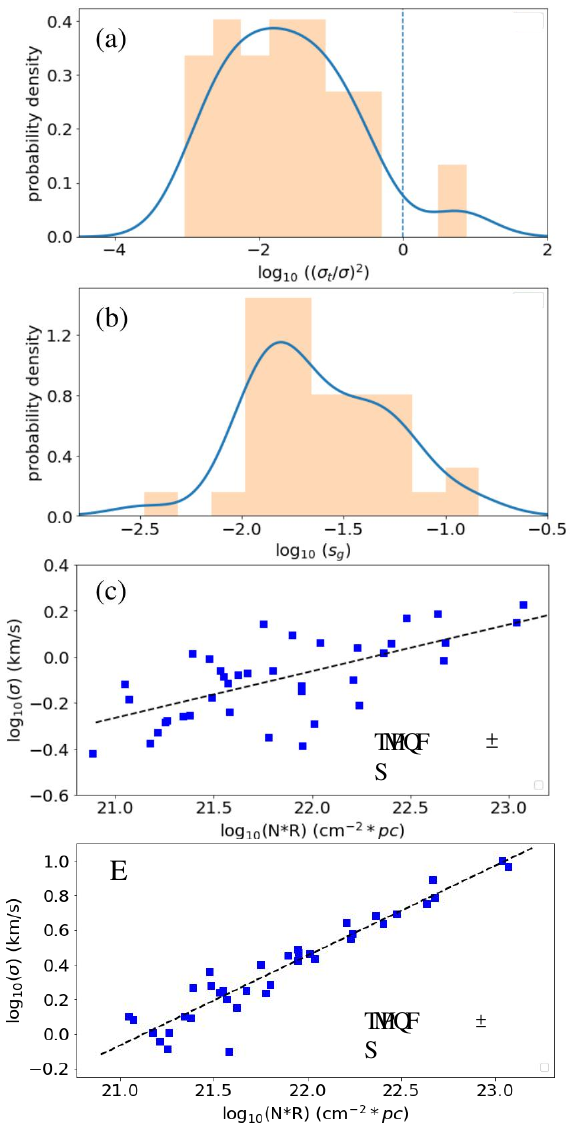}
\caption{(a) Distribution of the shear parameter $S_{g}$. (b) Distribution of the velocity dispersion ratio between $\sigma_{t}$ and $\sigma$. (c)  $\sigma-N*R$ relation, where $\sigma$ is the local gas velocity dispersion. (d)  $\sigma_{s}-N*R$ relation, where $\sigma_{s}=\sqrt{\sigma_{t}^{2}+\sigma^{2}}$. We note that “r” represents the Pearson correlation coefficient.}
\label{scaling}
\end{figure}

The very large velocity difference (velocity jump) around the protocluster in Fig.\ref{flow}, Fig.\ref{pv} and Fig.\ref{ppv} may indicate that the surrounding gas structures bear considerable shear and tides from the protocluster, apart from infall and rotation motions.
Fig.\ref{scaling}(c) shows the scaling relations between the velocity dispersion $\sigma$, the effective radius $R$, and the column density $N$ of all leaf structures, similar to \citet{Zhou2024-682} and \citet{Zhou2024-682-173}, 
the slope of the $\sigma_{s}-N*R$ relation is $\sim$0.2, which is much smaller than 0.5 and may indicate a pure free-fall gravitational collapse is slowing down.

To investigate the shear and tides from the protocluster,
here we analogy to the methods that are adopted on galaxy-cloud scales. For cloud-scale gas structures in the galaxy, as described in \citet{Hunter1998-493} and \citet{Dib2012-758}, the shear parameter $S_{g}$ is defined as
\begin{equation}
S_{g}=\frac{\Sigma_{\rm cri}}{\Sigma}.
\label{eq:6}
\end{equation}
We note that $\Sigma$ is the local gas surface density and 
$\Sigma_{\rm cri}$ is the critical surface density, which represents the minimal surface density needed to resist the shear. Following equation.7 of \citet{Dib2012-758},
\begin{equation}
\Sigma_{\rm cri}=\frac{A \sigma \ln(C)}{2\pi G} \, .
\label{eq:5}
\end{equation}The
$C$ represents the density contrast between leaf structures and their surroundings. The ratio between the average surface density derived from $^{13}$CO (3$-$2) emission in the FOV of the ALMA observation and the highest surface density of leaf structures is $\sim 100$, and thus we consider $C$=100.
We note that $\sigma$ is the local gas velocity dispersion and $A$ is the Oort constant, which measures the local shear level. For a spherical cloud of radius $r$, with a rotational velocity $V$ at a galactocentric distance $R_{G}$, $A$ can be calculated as 
\begin{equation}
A= 0.5 \left( \frac{V}{R_{G}} - \frac{\mid V(R_{G}+r) - V(R_{G}-r) \mid}{2r} \right) \, ,
\label{shear}
\end{equation}
in the context of molecular clouds, $V$ is determined by the rotation curve of the galaxy. In our case, for gas structures in the clump around the protocluster, $V$ was estimated by the Kepler rotation curve, as discussed in Sec.\ref{infall}. The $V = \sqrt{GM_{0}/R'}$, with $M_{0}$ being the mass of the protocluster and $R'$ being the distances between gas structures and the protocluster.
Shear is effective at tearing apart the gas structure if $S_{g}>1$ and it can be ineffective if $S_{g}<1$.

In a galactic context, the overall gas and stellar distribution can contribute to confine or disrupt the gas structures. For a cloud at a distance $R_{G}$ from the galactic center, the edge of the cloud from the galactic center is at a distance of $R_{G}$+$r$. The tidal acceleration in the cloud is $\sim T*r$, following \citet{Stark1978-225},
\begin{equation}
T=\frac{V^{2}}{R_{G}^{2}}-\frac{\partial}{\partial R_{G}}\left(\frac{V^{2}}{R_{G}}\right)  \, .
\label{tide}
\end{equation}
Inserting the Keplerian rotation curve, $T = 3GM_{0}/R'^{3}$, which is similar to equation.\ref{point}. Then the effective velocity dispersion from the tide for a structure is
$\sigma_{t}^2 \sim$ $T*R^2$, where $R$ is the effective radius of the structure.

In Fig.\ref{scaling}, $S_{g} << 1$ and
$\sigma_{t}^2/\sigma^2 << 1$; except for two structures adjacent to the protocluster, 
the results seem to deny the importance of the shear and tide from the protocluster. However, 
the analysis here is actually not entirely appropriate. For a leaf structure, it also bears the tides from other gas structures, not only the protocluster. 
In fact, it should be N-body tidal interactions; therefore, a complete tidal calculation would be more appropriate.


\subsection{Tidal field}\label{tide2}

\begin{figure*}
\centering
\includegraphics[width=0.9\textwidth]{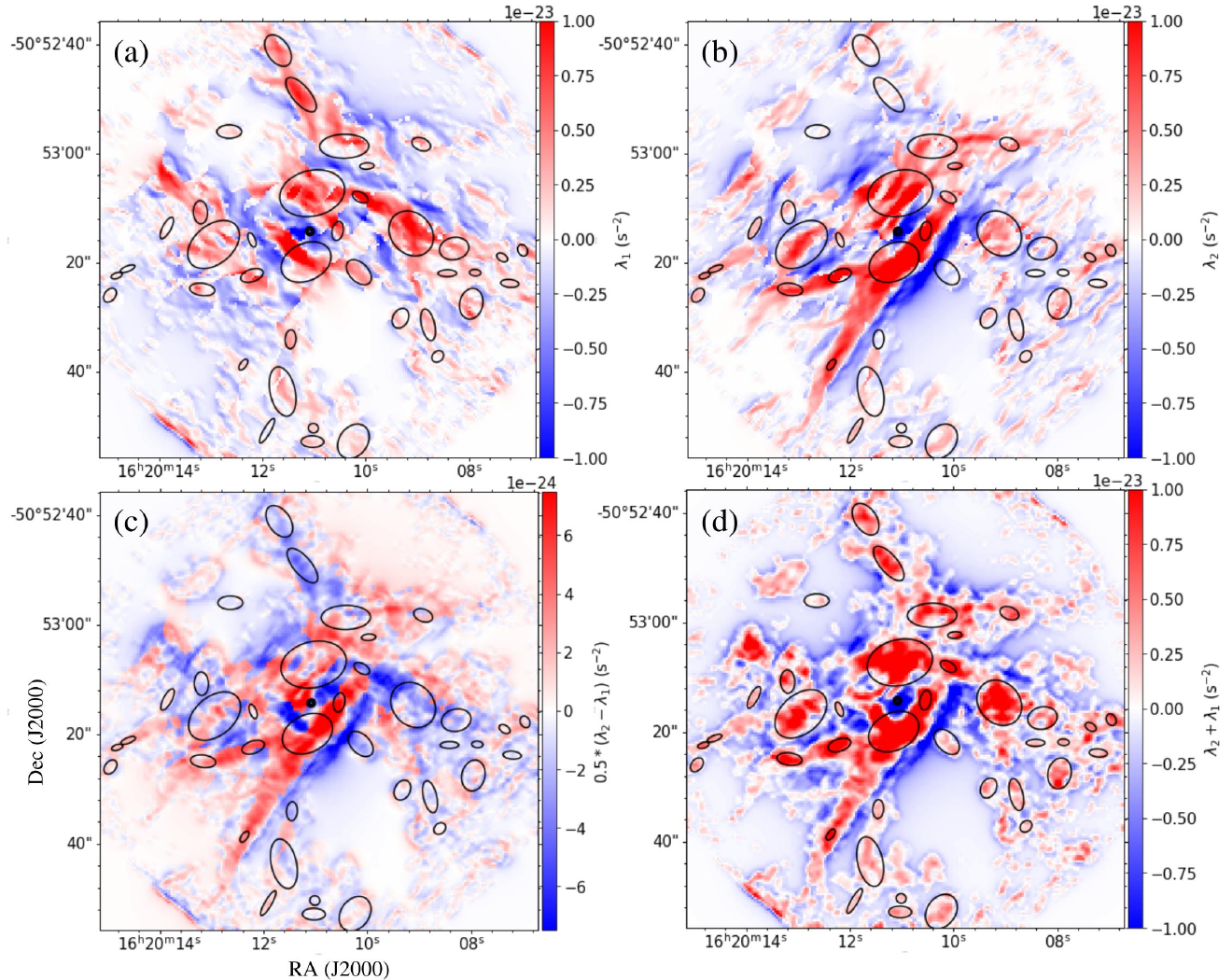}
\caption{Different components of the projected tidal field in the FOV of the ALMA observation.}
\label{tide}
\end{figure*}
To fully consider the tidal interactions,
we directly derived the tidal field according to the density distribution.
Only the projected surface density map is available in the observations. To align with other projected physical quantities derived from the observations, we also calculated the 2D projected tide.
We note that 3D calculations can refer to \citet{Ganguly2024} and \citet{Li2024-528}.
Calculating the gravitational potential field $\phi(x, y)$ based on the surface density map $\Sigma(x, y)$ has been discussed in \citet{Gong2011-729,He2023-526}. 
The tidal tensor described in Appendix.\ref{tide0}
has only two eigenvalues in the case of 2D, that is to say $\lambda_{i}$ (i=1,2). 
As described in Appendix.\ref{tide0}, 
the tidal tensor of a structure includes both self-gravity and external tides, that is, 
$\mathbf{T}= \mathbf{T}_{\mathrm{ext}} + \mathbf{T}_{\mathrm{int}}$.   
The decomposition of the tidal tensor is discussed in detail in Appendix.\ref{tide0}.
For three limiting cases, the external tides can be measured by $\lambda_{1}$, $\lambda_{2}$, and $(\lambda_{2}-\lambda_{1})/2$ and they are called T$_{\rm d,1}$, T$_{\rm d,2}$, and T$_{\rm d,3}$, respectively.
Fig.\ref{tide} displays different components of the tidal tensor. 

\subsection{Structure of G332.83-0.55}\label{sheet}

\begin{figure}
\centering
\includegraphics[width=0.48\textwidth]{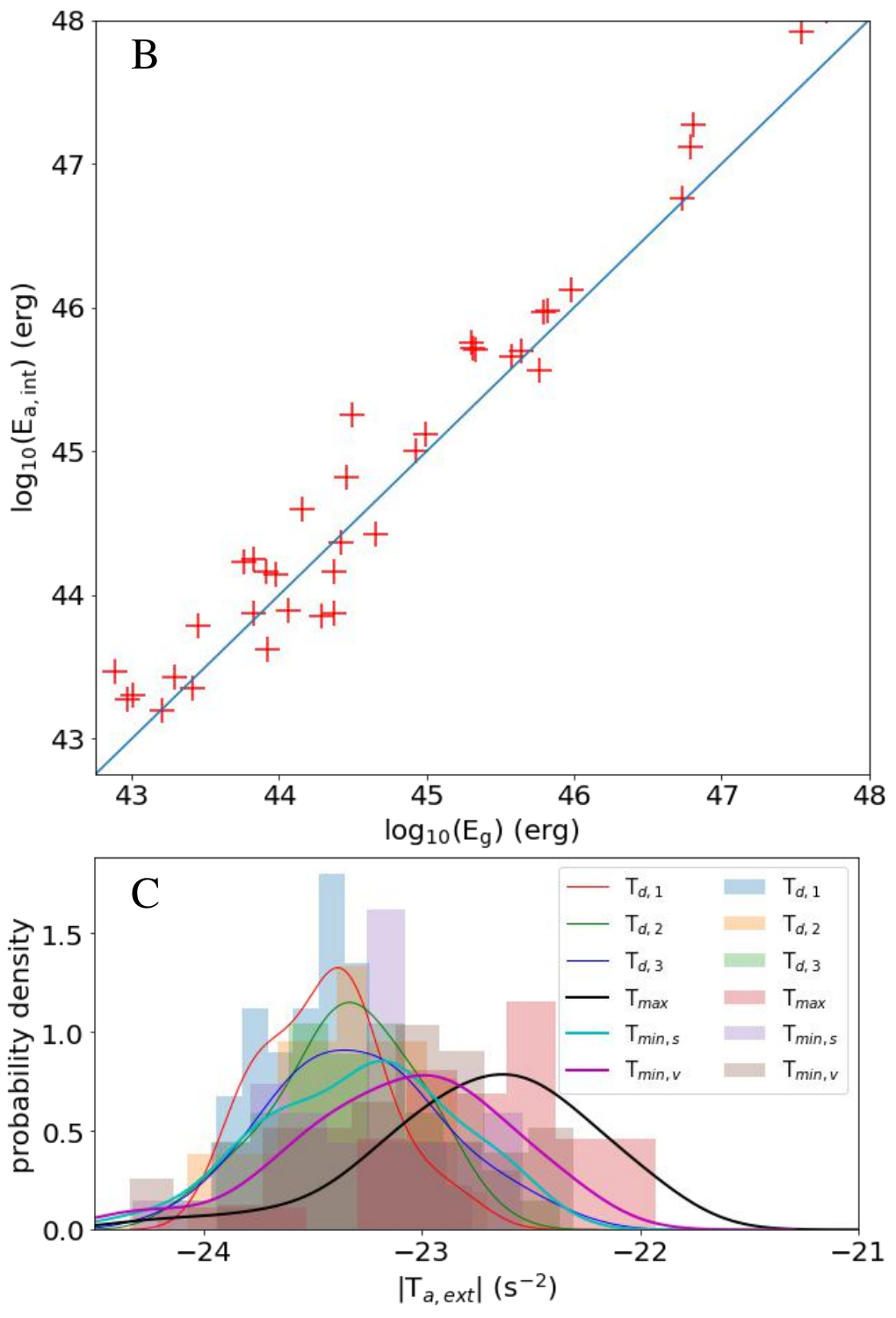}
\caption{(a) Comparison of $E_{\rm a,int}$ and $E_{\rm g}$ of dendrogram structures. (b) Distribution of the average tidal strength in each computation.}
\label{compare}
\end{figure}

As shown in Fig.1 of \citet{Zhou2023-676}, G332.83-0.55 is part of a large-scale shell structure. Thus, it may have a sheet structure. In combining Fig.\ref{map}(b) and Fig.\ref{ppv}(a), the single direction gas inflows on large scales also indicate that the structure on large scales should be approximately 2D, and should be face-on. Thus, we can observe the unidirectional converging flows. Close to the center, the structure becomes thicker and thicker. The redshifted and blueshifted components may originate from the forward and backward gas motions in the hub region. However, strong rotation motions on small scales may also explain their origin. 
The 3D and 2D structures of G332.83-0.55 may not be very different, given its limited thickness. Therefore, 2D-projected tidal field strength should be comparable to those for 3D ones, and equation.\ref{Et} derived from the 3D tidal tensor in Appendix.\ref{tide0} should still be effective for a 2D situation, which is confirmed in Fig.\ref{compare}(a). 

\subsection{Pixel-by-pixel computation}\label{pp}

Below we explain how we tested the above tidal tensor analysis using an original method.

\subsubsection{Maximum computation}\label{max}
\begin{figure}
\centering
\includegraphics[width=0.45\textwidth]{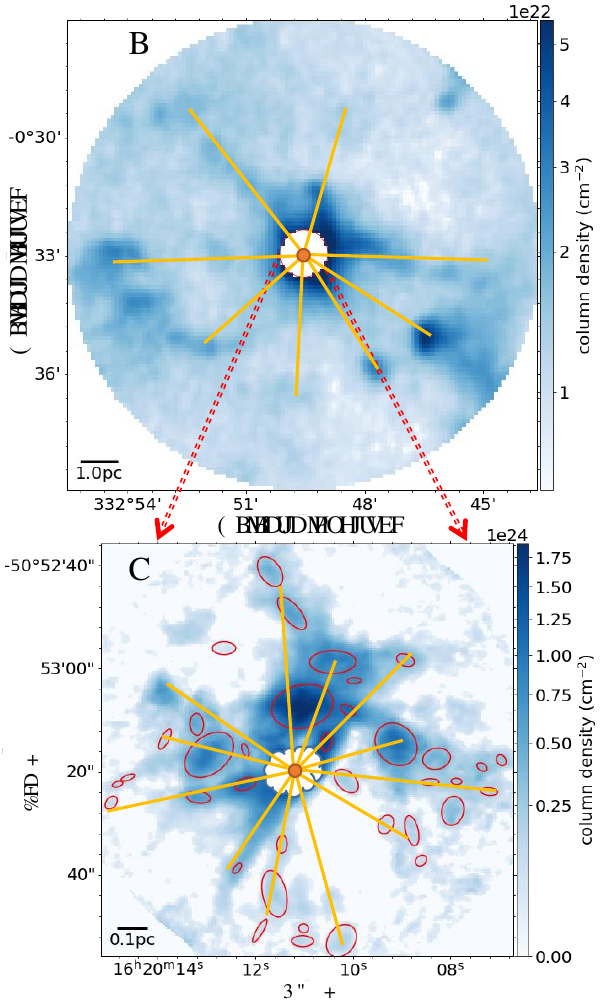}
\caption{Computed structure first masked from the density map. The external tides from all outside material sustained by a point in the computed structure was calculated pixel by pixel and is illustrated by orange lines. (a) Background representing the column density map derived from ATLASGAL+Planck 870 $\mu$m data. The material in the FOV of the ALMA observation (white region) was masked. (b) Background representing the column density map derived from H$^{13}$CO$^+$ J=1-0 and HCO$^+$ J=1-0 lines from the ALMA observation by the LTE analysis.}
\label{mask}
\end{figure}

\begin{figure}
\centering
\includegraphics[width=0.45\textwidth]{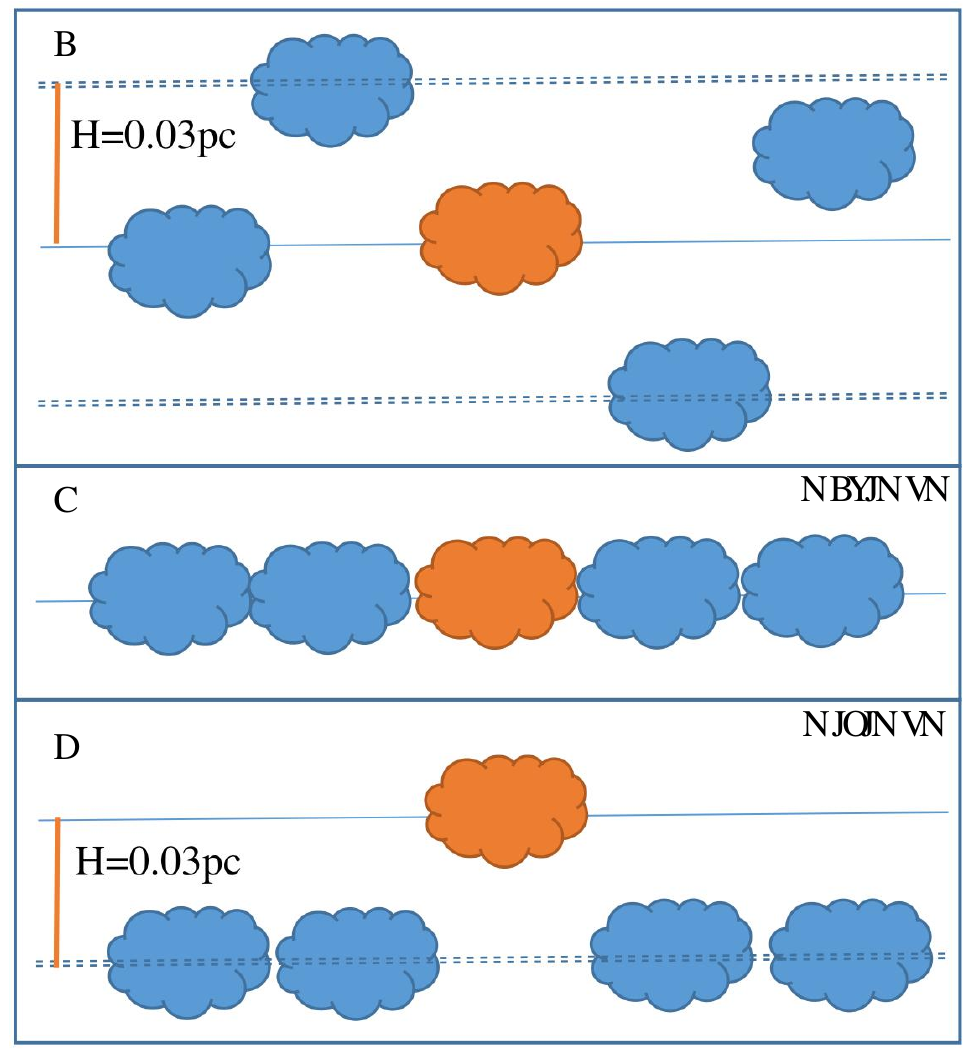}
\caption{Edge-on view of G332.83-0.55. The calculated gas structure is in red, and all external material apart from the calculated structure is in blue.
(I) The gas distribution in G332.83-0.55 can be everywhere. (II) In 2D computation, all molecular gases are assumed to be concentrated on the same plane. Thus, we obtained a maximal estimate for the tidal strength. (III) Taking a characteristic scale height of the molecular gas as 0.03 pc, all external material apart from the calculated structure were moved vertically to 0.03 pc away. Thus we obtained a minimal estimate of the tidal strength.}
\label{shift}
\end{figure}

For a structure A, located at a distance $R$ from the center of another structure B, of mass $M$, the tidal strength sustained by structure A due to the external gravity from structure B is
\begin{equation}
T \approx \frac{2GM}{R^{3}}.
\label{point}
\end{equation}
The total tidal strength at a pixel measures the total deformation due to the external tides at that point. 
A gas structure is neither a rigid body nor point mass, but it is a parcel of gas with malleability. 
Most gas structures are also not spherically symmetric, and their morphologies can be highly complex.
Although the external gravity from all directions may cancel each other out in a structure, the tides inside the structure due to the external gravity do not. Therefore, we used the scalar superposition to measure the overall tidal strength in the structure from the external gravity. Tidal tearing may be an important cause for the deformation and complex morphology of gas structures, as discussed in Sec.\ref{regulate}.

Using the mask of the dendrogram structure, we divided the density map into two parts, that is, the structure itself and all other materials apart from this structure. For the external tides from all outside material sustained by a point in structure A, a simple calculation involves using equation.\ref{point} to directly calculate the tidal strength at that point pixel by pixel, as illustrated in Fig.\ref{mask}(b). The average tidal strength of all the pixels in structure A can then represent the average tidal strength sustained by the structure.

\subsubsection{Minimal computation}\label{min}
As illustrated in Fig.\ref{mask}(b), the above computation is limited to a 2D plane. In equation.\ref{point}, the tidal strength is very sensitive to the distance. If the external structure and the calculated structure are not in the same plane, the 2D calculation underestimates the distance between them and thus overestimates the tidal strength (maximum computation, T$_{\rm max}$). However, we only have the 2D projection in the observation. The 3D structure of the molecular gas is unknown, but we can consider an extreme case. 
Using the average effective radius ($\sim$0.03 pc) of all leaf structures as the characteristic scale height of G332.83-0.55, all external material apart from the calculated structure was moved vertically to 0.03 pc away, as illustrated in Fig.\ref{shift}(c). This is unrealistic, given the continuity of the matter distribution in a cloud. As a limiting case, the estimated tidal strength should be smaller than the actual value (minimal computation, T$_{\rm min,s}$). 

The other minimal computation (T$_{\rm min,v}$) is the vector superposition of tides, which allows the tidal interactions to directly cancel each other out.
Using the calculated point in the target structure as the coordinate origin to build a\ 2D coordinate system, all external tides were decomposed to the coordinate axes. Then we calculated the overall tidal strength at the calculated point after performing vector superposition along each axis.

\subsubsection{Tide from the envelope}
As shown in Fig.\ref{flow1}, the dendrogram structures are embedded in a larger-scale gas environment. The material outside of the FOV of ALMA observation may also produce considerable external tides (T$_{\rm out}$) on small-scale dendrogram structures.
Using the H$_{2}$ column density derived from ATLASGAL+Planck 870 $\mu$m in \citet{Zhou2024-682}, the matter in the FOV of ALMA observation was first masked, as shown in Fig.\ref{mask}(a). Then we calculated the external tides at each pixel in the FOV. The averaged tidal strength of all pixels is $\sim$1.4 $\times 10^{-25}$ s$^{-2}$.
Compared with the tidal strength in Fig.\ref{compare}(b), T$_{\rm out}$ is very small and therefore can be ignored.

\subsubsection{Comparison}
Fig.\ref{compare}(b) shows that the components T$_{\rm d,1}$, T$_{\rm d,2}$, and T$_{\rm d,3}$ (whose derivation is detailed in Appendix.\ref{tide0}) are comparable.
Especially, they are
also comparable with T$_{\rm min,s}$ and T$_{\rm min,v}$. As expected, T$_{\rm max}$ is significantly larger than others.
For T$_{\rm d,1}$, T$_{\rm d,2}$, T$_{\rm d,3}$, T$_{\rm min,s}$, and T$_{\rm min,v}$, T$_{\rm min,v}$ is slightly larger than others and
the value of T$_{\rm min,s}$ is in the middle. As a compromise, we use the value of T$_{\rm min,s}$ to represent the external tides in the subsequent analysis.

\section{Discussion}\label{discussion}

\subsection{Tide-regulated gravitational collapse}\label{regulate}

\begin{figure}
\centering
\includegraphics[width=0.48\textwidth]{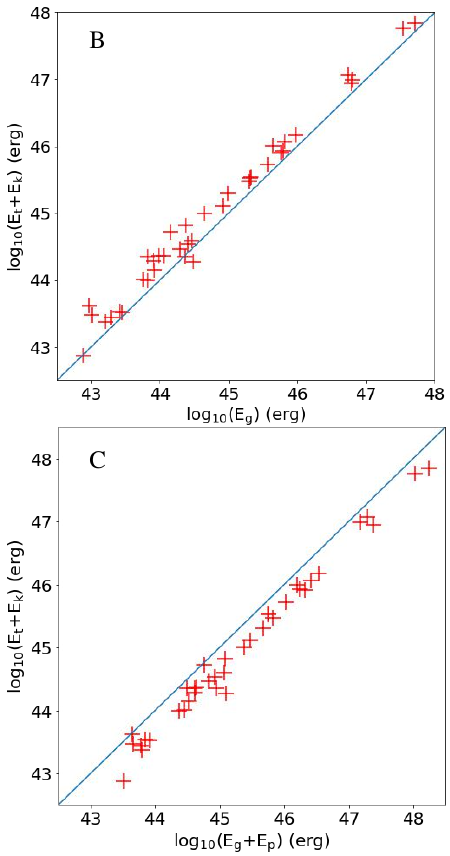}
\caption{Comparison between the structural energies, i.e., the gravitational potential energy E$_{\rm g}$, the kinetic energy E$_{\rm k}$, the tidal energy E$_{\rm t}$, and the external pressure energy E$_{\rm p}$.}
\label{energy}
\end{figure}

The average strength of the external tides T$_{\rm min,s}$ of a structure was used to measure the deformation of the structure due to the external gravity.
The effective velocity dispersion from the external tides of the structure is 
$\sigma_{t}^{2} \approx
T_{\rm min,s}*R^2$,
and the corresponding energy
\begin{equation}
E_{\rm t} \approx \frac{1}{2}
M *T_{\rm min,s}*R^2.
\label{tide}
\end{equation}
The gravitational potential energy $E_{\rm g}$ and internal kinetic energy $E_{\rm k}$ were calculated in the same way as presented in \citet{Zhou2024-682}.

The slowing down of a pure free-fall gravitational collapse as revealed by the shallow $\sigma_{\rm s}-N*R$ relation means that there are some physical processes working against gravitational collapse, such as tidal tearing from external gravity. 
The effect of the external tides on the gravitational collapse of gas structures can also be reflected in the scaling relation.
After adding the contribution of the tidal interactions to the observed velocity dispersion, the effective velocity dispersion $\sigma_{\rm s}=\sqrt{\sigma_{\rm t}^{2}+\sigma^{2}}$, then the $\sigma_{\rm s}-N*R$ relation was fitted again. 
As shown in Fig.\ref{scaling}(d), now the slope is close to 0.5, and the correlation also gets stronger. 
Thus, the deformation due to the external tides may effectively slow down the pure free-fall gravitational collapse of gas structures. 

However, as shown in Fig.\ref{energy}(a), the sum of E$_{\rm k}$ and E$_{\rm t}$ is notably larger than E$_{\rm g}$. 
The external tidal strength may be overestimated. 
The velocity dispersion is a manifestation of gas perturbations due to many physical processes, including gravitation collapse, tidal tearing, and turbulence.
Thus, E$_{\rm k}$ should be overestimated. Moreover,
when $E_{\rm k}+E_{\rm t}>E_{\rm g}$, it does not mean that the structure cannot be gravitationally bound.
A structure is deformed by external tides from the surrounding material, and at the same time, the surrounding material also provides the structure with a confining pressure. 

Previous studies have suggested that the external pressure provided by the larger-scale molecular gas might help to confine dense structures in molecular clouds \citep{Spitzer1978,McKee1989-345,Elmegreen1989-338,Ballesteros2006-372,Kirk2006-646,Dib2007-661,Lada2008-672,Pattle2015-450,Kirk2017-846,Li2020-896,Zhou2024-682}.
External pressure can have various origins; here we mainly consider the external pressure from the ambient cloud for each decomposed structure using
\begin{equation}
\label{evirp}
P_{\rm cl} = \pi G \bar{\Sigma}\Sigma_{r},
\end{equation}
where $P_{\rm cl}$ is the gas pressure, $\bar{\Sigma}$ is the mean surface density of the cloud, and $\Sigma_{\rm r}$ is the surface density at the location of each structure \citep{McKee1989-345,Kirk2017-846}.
The external pressure energy can be calculated as
\begin{equation}
\label{ep}
E_{\rm p} = -4\pi P_{\rm cl} R^{3}.
\end{equation}
As shown in Fig.\ref{energy}(b), after considering the external pressure,
$E_{\rm k}+E_{\rm t}<E_{\rm g}+E_{\rm p}$. The external tide tries to tear up the structure, but the external pressure on the structure prevents this process from occurring. 
The counterbalance between the external tide and external pressure hinders the gravitational collapse of the structure, which can also cause a pure free-fall gravitational collapse to slow down. 


Gravity can play multiple roles in cloud evolution. Inside dense regions, it drives fragmentation and collapse, and outside these regions, it can suppress the low-density gas from collapsing through extensive tidal forces and drive accretion \citep{Li2024-528}. 
This work confirms that tidal interactions can significantly influence the dynamics and kinematics of gas structures and regulate their gravitational collapse.
In Fig.14(c) and Fig.15(c) of \citet{Zhou2024-682}, in the G333 complex, for 3608 gas structures on cloud-clump scales, the slopes of the $\sigma-N*R$ relations are $\sim$0.19 and $\sim$0.26, respectively.
In the Fig.5(a) of \citet{Zhou2024-682-173}, for 2420 dense gas structures on clump-core scales in 64 ATOMS sources (clumps) at different evolutionary stages, the slope of the $\sigma-N*R$ relation is $\sim$0.27.
According to the above analysis,
these shallow slopes (significantly smaller than 0.5) may indicate extensive tide-regulated gravitational collapse from cloud to core scales, which will be discussed in detail in future work.

\subsection{Gravitational focusing}

In Fig.\ref{pv}(b) and Fig.\ref{ppv}(b), on large scales, the funnel structure is consistent with the schematic diagram in Fig.9 of \citet{Zhou2023-676}. However, it looks as though the bottom of the funnel splits into two parts in Fig.\ref{pv}(b) and Fig.\ref{ppv}(b).  
A velocity jump appears in the center comprised of the redshifted and blueshifted components, which may be attributed to the infalling and rotating motions around the protocluster. If we turn over the redshifted component and let it merge with the blueshifted component, then we obtain the funnel structure in Fig.9 of \citet{Zhou2023-676}. 

Similar to lights converging to the focus, now the gas inflows are converging to the gravitational center, that is, the hub region of G332.83-0.55.
Thus, there is a vivid "gravitational focusing" process, which also provides a clue as to the formation of the hub-filament structure. 
As shown in the Fig.1 of \citet{Zhou2023-676}, the entire G333 complex has a large-scale shell structure. G332.83-0.55 is a fragment on the shell, which should have a sheet configuration initially; for more details, readers can also refer to Sec.\ref{sheet}. Then the subsequent gravitational focusing process yields the hub-filament structure. In Fig.\ref{map}(b), the hub-filament structure is decoupling from the surrounding gas environment due to gravitational collapse.

\section{Summary}

1. In high-resolution APEX/LAsMA $^{12}$CO and $^{13}$CO (3-2) observations,
G332.83-0.55 has a clear hub-filament morphology at $\sim$10
pc, and its hub corresponds to the ATLASGAL clump AGAL332.826-00.549. We zoomed in on the hub region of G332.83-0.55 using higher-resolution ALMA HCO$^{+}$ (1-0) and H$^{13}$CO$^{+}$ (1-0) observations, and found that it still presents a hub-filament structure. Thus, G332.83-0.55 contains at least two levels of hub-filament structures. High-resolution single-dish and ALMA observations provide the opportunity to trace gas inflows from cloud to core scales.

2. G332.83-0.5 is collapsing with rotation. Around the gravitational center (i.e., the protocluster), both the velocity fields traced by LAsMA and ALMA observations show velocity jumps comprised of redshifted and blueshifted peaks. According to the velocity difference and the separation between blueshifted and redshifted peaks, we roughly estimated the rotation velocity $V_{\rm rotation,ALMA} \approx$3.29 km/s and the centrifugal barrier $r_{\rm c, ALMA} \approx$ 0.21 pc. Then the mass of the gravitational center (i.e., the protocluster) was estimated to be $\approx$264 M$_{\odot}$, which agrees with the mass estimated by 3mm dust emission.

3. We directly identified local dense gas structures based on the 2D integrated intensity (Moment 0) map of H$^{13}$CO$^+$ J=1-0 emission using the dendrogram algorithm, and then extracted the average spectra to fit the velocity dispersion. The slope of the $\sigma-N*R$ relation for leaf structures is $\sim$0.2, which is much smaller than 0.5 of the pure free-fall gravitational collapse.\ This may indicate that a pure free-fall gravitational collapse is slowing down.

4. We investigated the effect of shear and tides from the protocluster on the surrounding local dense gas structures. The results seem to deny the importance of shear and tides from the protocluster. However, for a gas structure, it also bears the tides from other gas structures and the general gas distribution, not only the protocluster. 
To fully consider the tidal interactions,
we derived the tide field according to the density distribution, and then used the average strength of the external tides of a structure to measure the total tidal interactions on the structure. For comparison, we also adopted an original pixel-by-pixel computation to estimate the average tidal strength of each structure. Both methods give comparable calculation results.

5. After considering the tidal interactions for all structures, the slope of the $\sigma-N*R$ relation is close to 0.5, and the correlation also gets stronger. Thus, the deformation due to the external tides can effectively slow down the pure free-fall gravitational collapse of gas structures. 
The external tide tries to tear up the structure, but the external pressure on the structure prevents this process from occurring. 
The counterbalance between an external tide and external pressure hinders the free-fall gravitational collapse of the structure, which can also cause a pure free-fall gravitational collapse to slow down. These mechanisms can be called "tide-regulated gravitational collapse."

\begin{acknowledgements}


This publication is based on data acquired with the Atacama Pathfinder Experiment (APEX) under programme ID M-0109.F-9514A-2022. APEX has been a collaboration between the Max-Planck-Institut für Radioastronomie, the European Southern Observatory, and the Onsala Space Observatory.
This research made use of astrodendro, a Python package to compute dendrograms of Astronomical data (http://www.dendrograms.org/). 
This paper makes use of the following ALMA data: ADS/JAO.ALMA$\sharp$2019.1.00685.S. ALMA is a partnership of ESO (representing its member states), NSF (USA) and NINS (Japan), together with NRC (Canada), MOST and ASIAA (Taiwan), and KASI (Republic of Korea), in cooperation with the Republic of Chile. The Joint ALMA Observatory is operated by ESO, 

\end{acknowledgements}

\bibliography{ref}
\bibliographystyle{aa}


\begin{appendix}

\section{Tidal tensor}\label{tide0}

For a gravitational potential field $\Phi$, the tidal tensor $\mathbf{T}$ is defined as
\begin{equation}
    \mathrm{T}_{ij} = \partial_i \partial_j \Phi.
\end{equation}
The eigenvalues of the tidal tensor contain information regarding the extent of deformation (compression and disruption) experienced at a specific point due to the local gravitational field.
The diagonalized tidal tensor is given as 
\begin{gather}
    \mathbf{T}
    =
    \begin{bmatrix}
    \lambda_1 &0&0\\
    0&\lambda_2&0\\
    0&0&\lambda_3
    \end{bmatrix}.
\end{gather}
The sign of the eigenvalues ($\lambda_1, \lambda_2,$ and $   \lambda_3$) indicates whether the given mode is compressive ($\lambda_i > 0$) or disruptive ($\lambda_i < 0$), while the magnitude represents the strength of the respective compressive and disruptive mode. 
The trace of the tidal tensor $\mathrm{Tr(\mathbf{T})}=\sum_{i=1}^{3} \lambda_i$ contains the local density information in the Poisson equation:
\begin{equation}
    \mathrm{Tr(\mathbf{T})} = \nabla^2 \Phi = 4\pi G \rho.
    \label{trace}
\end{equation}
This indicates that the trace is zero when the location where the tidal tensor is assessed lies outside the distribution of mass.

As was done in \citet{Ganguly2024}, the tidal tensor can be split into three parts: the tidal tensor due to only the matter inside the structure, $\mathbf{T}_{\mathrm{int}}$, and into a second component due only to the matter distribution external to the structure, $\mathbf{T}_{\mathrm{ext}}$. Their sum, due to the entire matter distribution,
\begin{equation}
\mathbf{T} =\mathbf{T}_{\mathrm{int}}+\mathbf{T}_{\mathrm{ext}}.
\label{3t}
\end{equation}
Here, $\mathrm{\mathbf{T}}$ represents the net deformation due to both the structure itself and its surroundings. We note that $\mathrm{\mathbf{T}_{ext}}$ represents the deformation introduced solely due to the external materials. The trace of the three parts are the following:
\begin{gather}\label{eq:trace1}
    \mathrm{Tr(\mathbf{T})} = 4\pi G \rho_{\mathrm{a}}\\
    \mathrm{Tr(\mathbf{T}_{int})} = 4\pi G \rho_{\mathrm{a}}\\
    \mathrm{Tr(\mathbf{T}_{ext})} = 0,
    \label{trace3}
\end{gather}
where $\rho_{\mathrm{a}}$ is the average density of the structure with radius $R$ and total mass $M$, $\rho_{\mathrm{a}}= M/(4/3 \pi R^{3})$. 

According to
$\mathrm{Tr(\mathbf{T}_{int})} = 4\pi G \rho_{\mathrm{a}}$,
the corresponding energy can be estimated by 
\begin{equation}
E_{\rm a,int} \approx \frac{1}{2}
M *\mathrm{Tr(\mathbf{T}_{int})}* R^{2} \approx \frac{GM^{2}}{R}.
\label{Et}
\end{equation}
Thus, the tidal tensor $\mathbf{T}_{\mathrm{int}}$ represents the self-gravity of the structure. If we want to estimate the tidal interactions from all external materials on the structure, we should focus on $\mathbf{T}_{\mathrm{ext}}$. 


The trace in equation.\ref{trace3} indicates that
$\mathrm{\mathbf{T}_{ext}}$ due to only matter distribution external to the structure must contain both compressive and disruptive modes, corresponding to positive and negative $\lambda_i$, respectively. In contrast, $\mathrm{\mathbf{T}_{int}}$ due to only the matter inside the structure and $\mathrm{\mathbf{T}}$ due to the entire matter distribution must contain at least one compressive mode, but it might also be fully compressive. These rules are also applicable for 2D projected tidal tensor. 
According to equation.\ref{3t},
the 2D tidal tensor can be decomposed as
\begin{equation}
\begin{bmatrix}
    \lambda_1 &0\\
    0&\lambda_2
    \end{bmatrix} =  \begin{bmatrix}
    -A_{0} &0\\
    0&A_{0}
    \end{bmatrix} +  \begin{bmatrix}
    A_{1} &0\\
    0&A_{2}
    \end{bmatrix};\end{equation}
assuming $A_{2}=k*A_{1}$, we have $A_{1}=\frac{\lambda_1+\lambda_2}{1+k}$, $A_{2}=\frac{k(\lambda_1+\lambda_2)}{1+k}$, and 
$A_{0}=\frac{\lambda_1+\lambda_2}{1+k}-\lambda_1$. There are three cases:

(1) When $k=1$, $A_{1}=A_{2}=\frac{\lambda_1+\lambda_2}{2}$, $A_{0}=\frac{\lambda_2-\lambda_1}{2}$ (Fig.\ref{tide}(c)).

(2) When $k<<1$, $A_{1} \approx \lambda_1+\lambda_2$, $A_{2} \approx 0$,  $A_{0} \approx \lambda_2$ (Fig.\ref{tide}(b)).

(3) When $k>>1$, $A_{1} \approx 0$, $A_{2} \approx \lambda_1+\lambda_2$,  $A_{0} \approx -\lambda_1$ (Fig.\ref{tide}(a)).

We estimated the averaged strength of the external tidal field for each structure according to Fig.\ref{tide}(a), (b) and (c), that is, the averaged value of $|A_{0, a}|$ in each structure. As shown in Fig.\ref{compare}(b), the three cases give similar estimates of the averaged strength of the external tidal field.

\end{appendix}

\end{document}